\documentclass[10pt,preprint]{aastex}
\newcommand\msun{\ensuremath{\mathcal{M}_\sun}}
\newcommand\R{\mathit{\Gamma}}

\begin{document}
\shorttitle{PA-99-N2}
\shortauthors{the POINT-AGAPE collaboration}
\title{The Anomaly in the Candidate Microlensing Event PA-99-N2}
\author{
Jin H. An,\altaffilmark{1}
N. W. Evans,\altaffilmark{1}
E. Kerins,\altaffilmark{2}
P. Baillon,\altaffilmark{3}
S. Calchi~Novati,\altaffilmark{4}
B. J. Carr,\altaffilmark{5}\\
M. Cr\'ez\'e,\altaffilmark{6,7}
Y. Giraud-H\'eraud,\altaffilmark{6}
A. Gould,\altaffilmark{8}
P. Hewett,\altaffilmark{1}
Ph. Jetzer,\altaffilmark{4}
J. Kaplan,\altaffilmark{6}\\
S. Paulin-Henriksson,\altaffilmark{6}
S. J. Smartt,\altaffilmark{1}
Y. Tsapras,\altaffilmark{5}
and D. Valls-Gabaud\altaffilmark{9}
\\(The POINT-AGAPE Collaboration)}
\altaffiltext{1}
{Institute of Astronomy, University of Cambridge,
Madingley Road, Cambridge CB3 0HA, UK}
\altaffiltext{2}
{Astrophysics Research Institute, Liverpool John Moores University,
12 Quays House, Egerton Wharf, Birkenhead CH41 1LD, UK}
\altaffiltext{3}
{CERN, CH-1211 Gen\`eve 23, Switzerland}
\altaffiltext{4}
{Institut f\"ur Theoretische Physik, Universit\"at Z\"urich,
Winterthurerstrasse 190, CH-8057 Z\"urich, Switzerland}
\altaffiltext{5}
{Astronomy Unit, School of Mathematical Sciences, Queen Mary, 
University of London, Mile End Road, London E1 4NS, UK}
\altaffiltext{6}
{Laboratoire de Physique Corpusculaire et Cosmologie,
Coll\`ege de France, 11 Place Marcelin Berthelot, F-75231 Paris,
France}
\altaffiltext{7}
{Universit\'e Bretagne-Sud, Campus de Tohannic, BP 573,
F-56017 Vannes Cedex, France}
\altaffiltext{8}
{Department of Astronomy, Ohio State University,
140 West 18th Avenue, Columbus, OH 43210, USA}
\altaffiltext{9}
{Laboratoire d'Astrophysique, Observatoire Midi-Pyr\'en\'ees,
14 Avenue Edouard Belin, F-31400 Toulouse, France}
%
\begin{abstract}
The lightcurve of PA-99-N2, one of the recently announced microlensing
candidates towards M31, shows small deviations from the standard
Paczy\'nski form. We explore a number of possible explanations,
including correlations with the seeing, the parallax effect and a
binary lens. We find that the observations are consistent with an
unresolved RGB or AGB star in M31 being microlensed by a binary lens.
We find that the best fit binary lens mass ratio is $\sim 1.2\times
10^{-2}$, which is one of most extreme values found for a binary lens
so far. If both the source and lens lie in the M31 disk, then the
standard M31 model predicts the probable mass range of the system to
be 0.02-3.6 \msun\ (95\% confidence limit). In this scenario, the mass
of the secondary component is therefore likely to be below the
hydrogen-burning limit. On the other hand, if a compact halo object in
M31 is lensing a disk or spheroid source, then the total lens mass is
likely to lie between 0.09-32 \msun, which is consistent with the
primary being a stellar remnant and the secondary a low mass star or
brown dwarf. The optical depth (or alternatively the differential
rate) along the line of sight toward the event indicates that a halo
lens is more likely than a stellar lens provided that dark compact
objects comprise no less than 15\% (or 5\%) of haloes.
\end{abstract}


\section{Introduction}

The POINT-AGAPE\footnote{Pixel-lensing Observations on the Isaac
Newton Telescope - Andromeda Galaxy Amplified Pixels Experiment}
collaboration is a part of a wider group of investigators monitoring
M31 \citep[e.g.,][]{An97,CN02,CN03} for the purpose of discovering
microlensing events, and, in particular, gradients in the rate of
events that should be induced by the high inclination of the M31 disk
\citep{Cr92,Ba93}, so as to infer the presence of dark compact objects
in the M31 halo. Recently, the POINT-AGAPE collaboration announced the
discovery of four high signal-to-noise ratio candidate events in the
data taken during 1999 and 2000 seasons\citep[ see also
\citealt{PA99N1} and \citealt{PA00S4}]{PH03}. The
MEGA\footnote{Microlensing Exploration of the Galaxy and Andromeda}
collaboration has also reported provisional events \citep{JDJ}.

This paper studies one of the candidate events, PA-99-N2, which was
reported to exhibit an anomaly in its lightcurve. Provided that this
is not due to any small but systematic photometric error, some
physical mechanism behind this behavior must be sought. This is a
productive avenue to explore as it is possible to infer some
properties of the lens or source from the deviations from the standard
Paczy\'nski form. Here, we show that the observed deviations are
consistent with being due to the close approach to a caustic in a
binary lens. We calculate the parameters of the best fit binary
solutions and discuss their implications.

\section{The Anomaly}
\label{data}

\subsection{The Data}

The POINT-AGAPE collaboration has acquired M31 imaging data for three
seasons using the Wide Field Camera on the 2.5 m Isaac Newton
Telescope located at La Palma, the Canary Islands, Spain. The full
analysis of this data set is in progress. \citet{PH03} reported four
microlensing candidate events in the first two seasons of data with
full-width half-maximum timescales less than 25 days and with flux
variations greater than that of a 21st magnitude star (in $R$
band). Among them is PA-99-N2, which is the second event identified in
the northern field with a maximum during the 1999 season. It is a very
bright event (at maximum) with high signal-to-noise ratio, lying at
22\arcmin03\arcsec\ ($\sim$5 kpc in projected distance) from the
center of M31. Other characteristics are listed in Table~1 of
\citet{PH03}. The event was observed in three bands, close to the
Sloan $g'$, $r'$ and $i'$ filters. As are all POINT-AGAPE
observations, the data have been reduced and analyzed using the
superpixel method \citep{CN02,PH03}. The transformation to
(Johnson/Cousins) $VRI$ is derived from the observations of
\citet{Hi94}.

\begin{figure}
\plotone{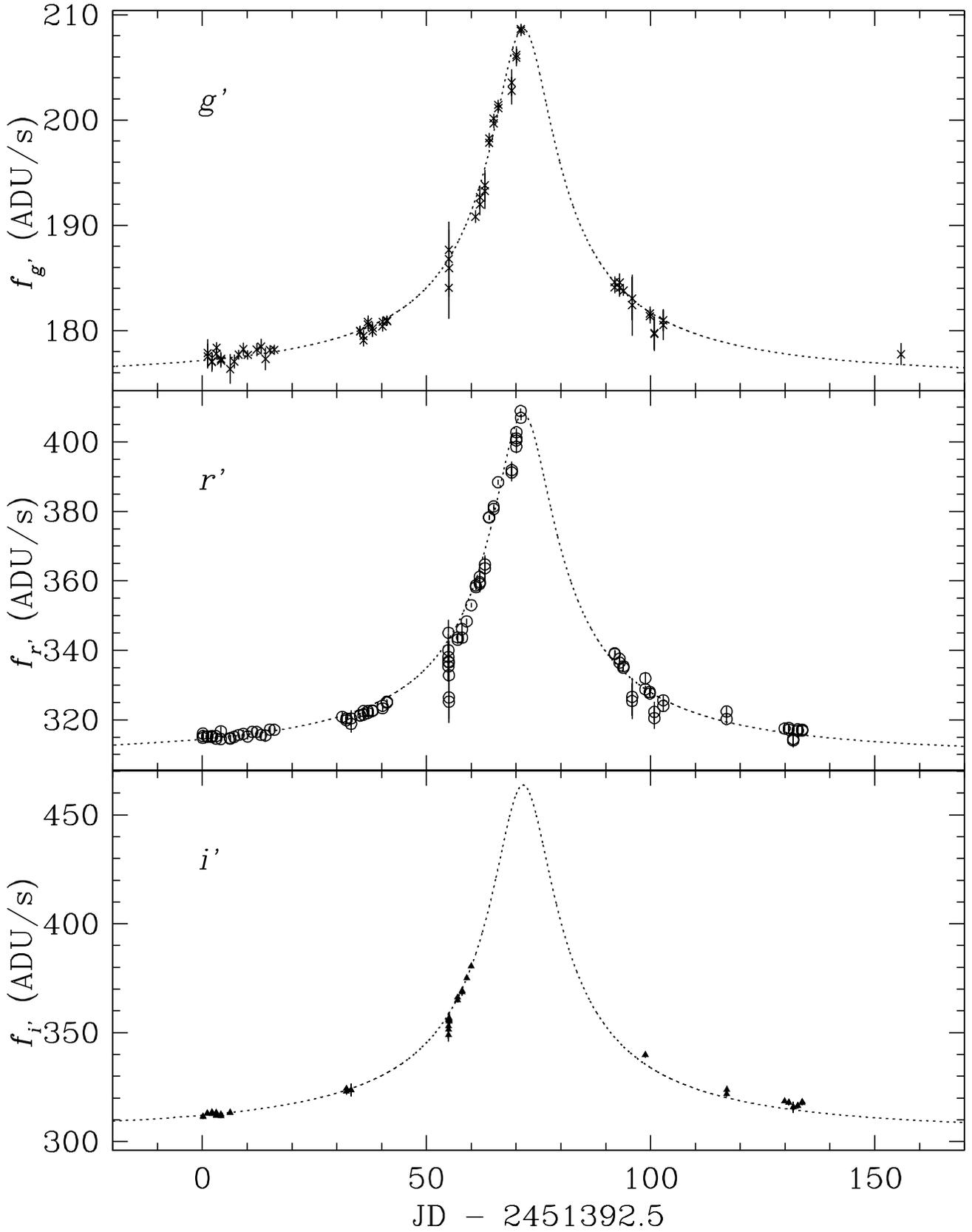}
\caption{\label{lightcurve}
Lightcurve of PA-99-N2 in the $r'$, $g'$ and $i'$ bands for the 1999
season. The dotted line shows the best fit Paczy\'nski lightcurve which
assumes an isolated point-mass lens. Crosses in the top panel are for
$g'$, open circles in the middle panel are for $r'$, and filled
triangles in the bottom panel are for $i'$ datapoints.
}\end{figure}

\begin{figure}
\plotone{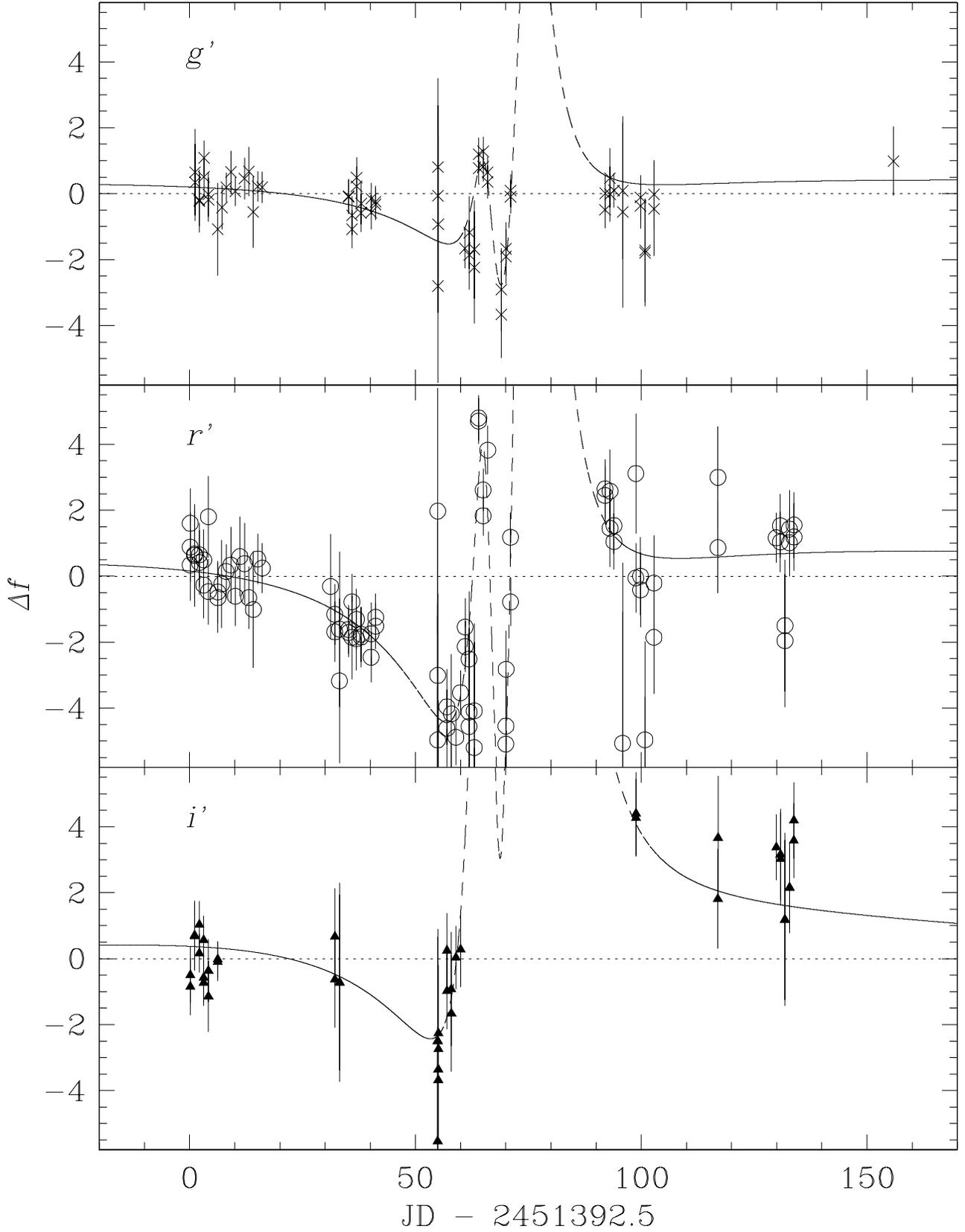}
\caption{\label{binresiduals}
Residuals from a standard Paczy\'nski fit. Symbols are the same as
Fig.~\ref{lightcurve}. The curve shows the best fit binary lens model
(C1).
}\end{figure}

\begin{figure}
\plotone{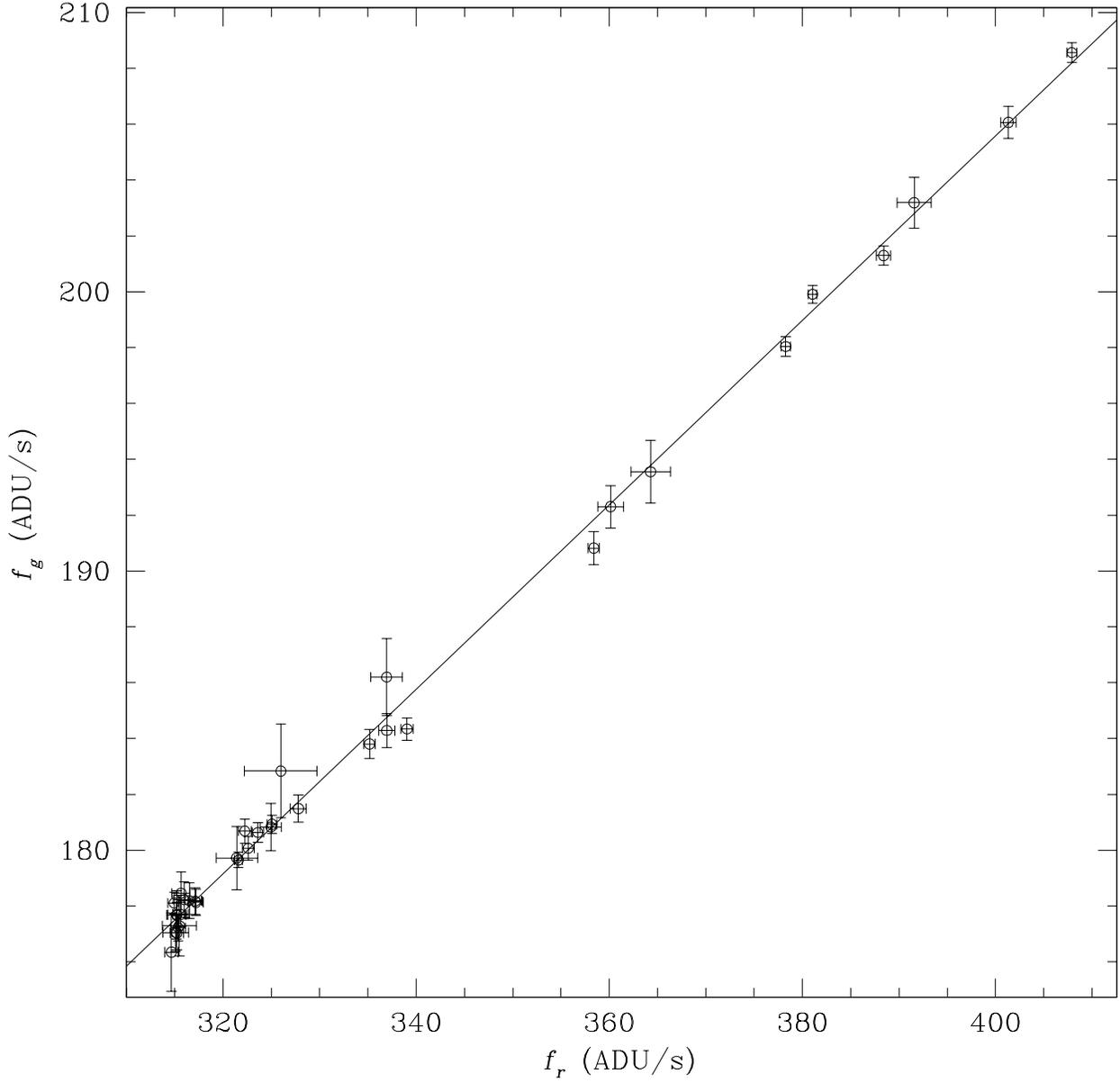}
\caption{\label{regression} 
Daily averaged $g'$ band flux plotted against $r'$ band flux observed
at the same night. Also shown as a solid line is the best fit linear
regression line. Note that the slope of this regression line is
related to the instrumental color $g'_{\rm I}-r'_{\rm I}=-2.5\log a$.
}\end{figure}

Figure~\ref{lightcurve} shows the lightcurve of PA-99-N2 resulting
from the analysis based on the superpixel method in all three bands
for the 1999 season. In Figure~\ref{binresiduals}, we also present the
residuals from the best fit standard Paczy\'nski lightcurve (PL).
There are 65 datapoints in $g'$, 102 in $r'$ and 45 in $i'$. The
lightcurve is flat during the 2000 and 2001 seasons. Visual inspection
of the lightcurve shows that there are systematic deviations around
JD$'$ 65, which have the same pattern in the $g'$ and $r'$ bands at
the same times (there were no $i'$ observations made at the time of
anomaly). Henceforth, JD$'$ denotes the Julian Date minus 2451392.5.
In Figure~\ref{regression}, we plot the daily averaged $g'$
datapoints against the daily averaged $r'$ datapoints whenever those
two band observations are available for the same night. The excellent
linear correlation between them confirms the achromaticity of the
variable fluxes including the anomalies in the lightcurve. The
standard Paczy\'nski fit to daily averaged datapoints gives a $\chi^2$
per degree of freedom of 3.1 \citep{PH03}, which indicates that the
deviations are statistically significant.

\begin{figure}
\plotone{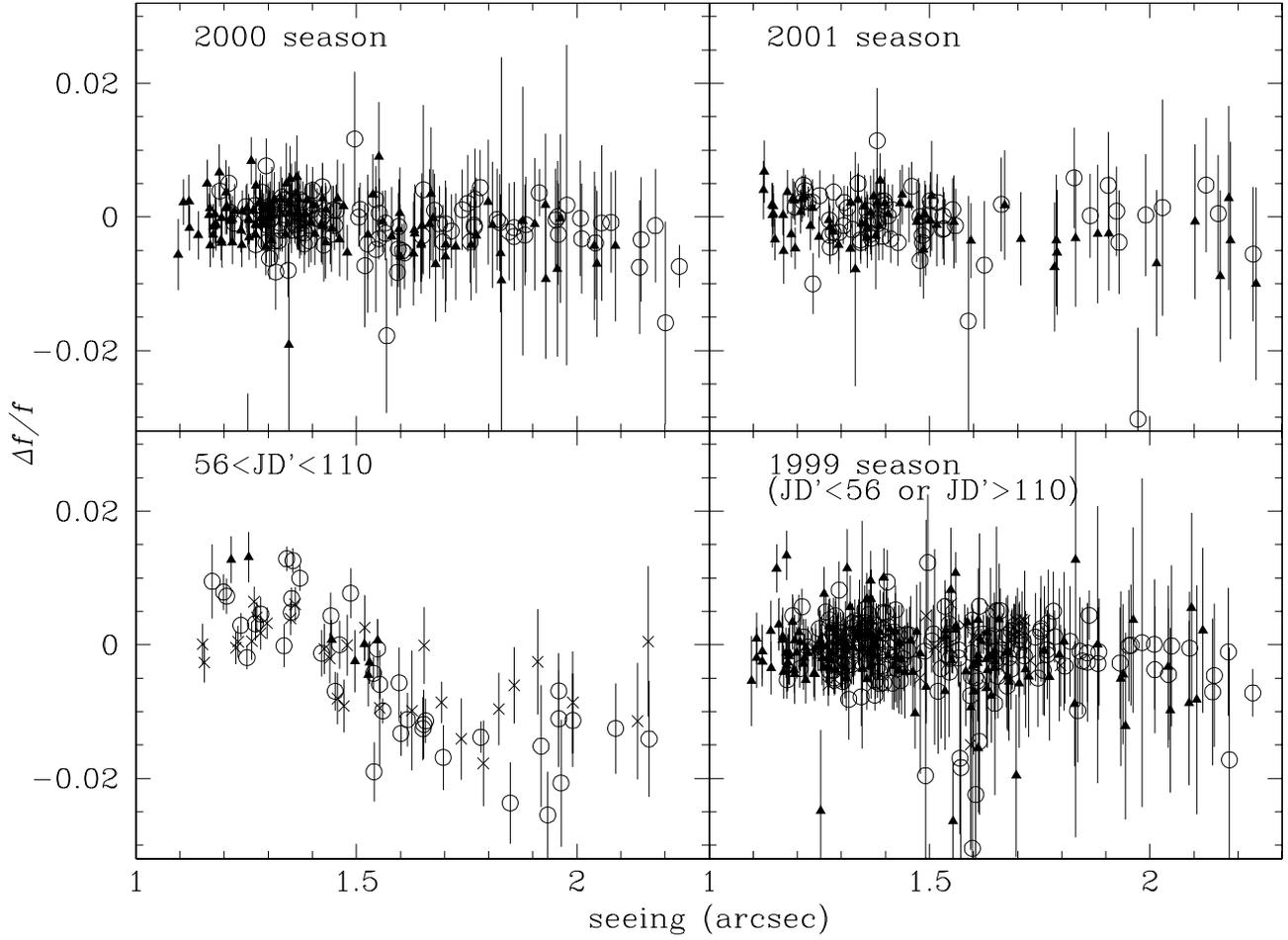}
\caption{\label{correlations}
Relative residuals versus seeing. Symbols are the same as
Fig.~\ref{lightcurve}. For 1999 season, residuals are from the best
fit standard Paczy\'nski curve. For 2000 and 2001 seasons, residuals
are with respect to flat baseline. 
}\end{figure}

The photometric recalibration \citep{CN02,PH03} of the superpixel
method may lead to some concern that the deviations are an artifact of
the reduction pipeline. We find a statistical correlation between the
seeing and residuals of the superpixel flux with respect to the best
fit Paczy\'nski curve for the datapoints taken during the deviation
(Fig.~\ref{correlations}). This raises the concern that the anomaly
may be just an artifact due to seeing variations, and that the
superpixel method does not make the proper recalibration. Although we
emphasize that there exists no such correlation for all other
datapoints (Fig.~\ref{correlations}), it is still worrisome as the
superpixel method was not devised for the photometry of bright
resolved sources, for which the apparent correlations take place.

\begin{figure}
\plotone{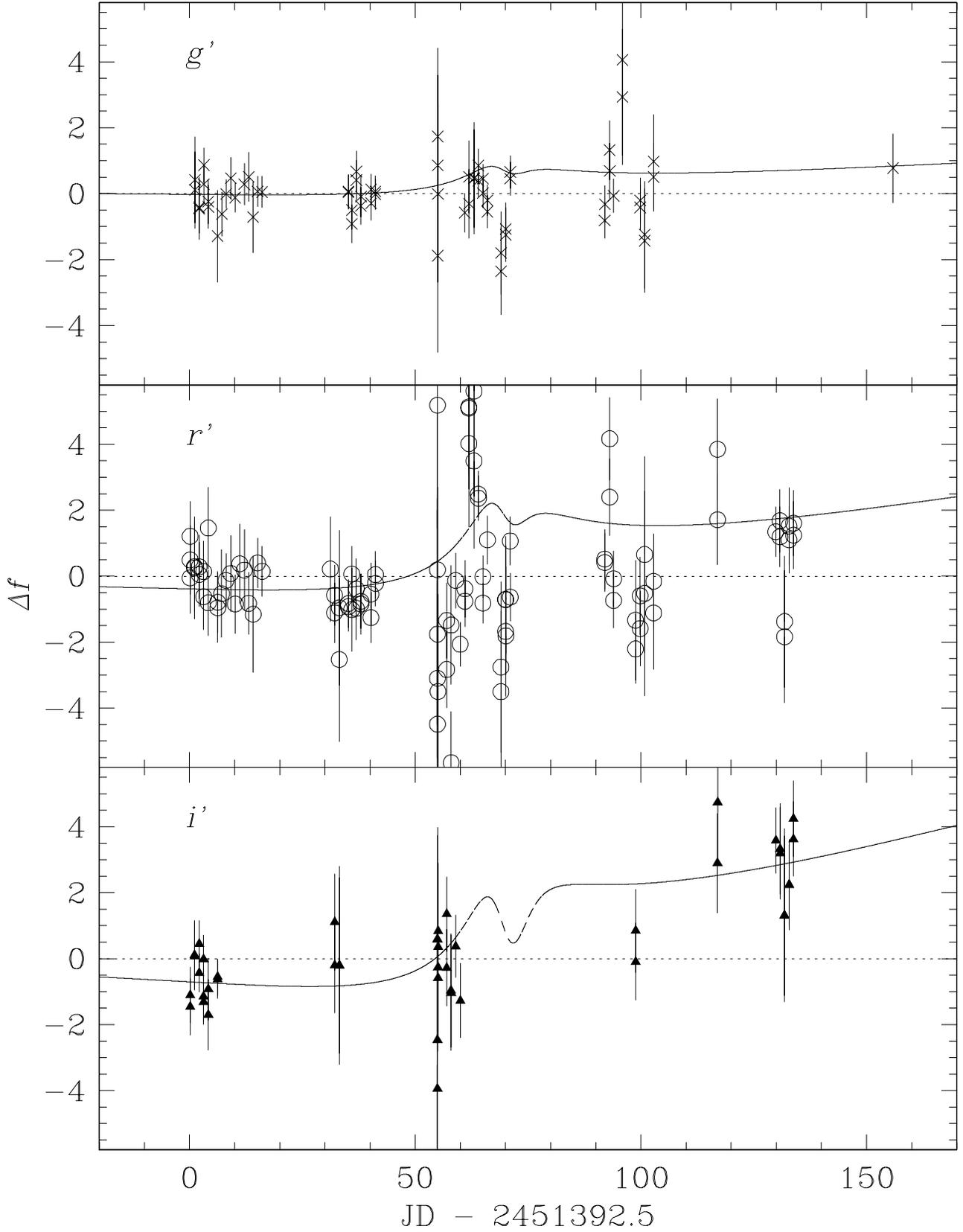}
\caption{\label{newcurves}
Residuals for the Paczy\'nski curve fit with seeing correlation given
by eq.~(\ref{eq:seevar}). Symbols are the same as
Fig.~\ref{binresiduals}. The curve shows the pattern of residuals
expected for the parallax model (XPN) given in the
Appendix~\ref{appb}.
}\end{figure}

To examine the hypothesis that the deviations can be ascribed solely
to seeing variations, we model the reported superpixel flux $f$, when
the source is resolved, by
\begin{equation}
f = f_{\rm true}[1 + \eta(\theta-\theta_0)]
\label{eq:seevar}
\end{equation}
where $f_{\rm true}$ is the true flux, $\theta$ is the seeing, and
$\eta$ and $\theta_0$ are fit parameters, while $f=f_{\rm true}$
(i.e., $\eta=0$) for all the other datapoints. Note that this provides
a simple way of describing the correlation of flux with seeing found
in Figure~\ref{correlations}. When we fit for $\eta$ and $\theta_0$ by
requiring the behavior of the true flux to be described by a
Paczy\'nski curve, this results in significant detections of nonzero
$\eta$'s ($\eta_{g'}=-0.020\pm0.003$, $\eta_{r'}=-0.043\pm0.003$,
$\eta_{i'}= -0.06\pm0.01$) with an overall better fit to the
Pacz\'nski curve, as the $\chi^2$ decreases from 824.7 (700.9) to
438.1 (321.1).  Here, the first number is from the data for 1999-2000
seasons (488 datapoints), while the second number is from the 1999
season data only (212 datapoints). Figure~\ref{newcurves} shows the
residuals for this fit. While there is no visible anomaly any more
around JD$'$ 65, close inspection reveals that the lightcurve is still
not entirely consistent with Paczy\'nski curve; the datapoints lie
systematically below the rising parts of the model lightcurve (in
particular, $55<\mbox{JD}'<67$) and the datapoints lie above the model
in the falling side wing.\footnote{The pattern of the deviations is
reminiscent of the parallax effect and this possibility is explored in
Appendix~\ref{appb}.} In other words, although there is a correlation
between flux residuals and seeing, this by itself is not enough to
explain the shape of the lightcurve.

\begin{figure}
\plotone{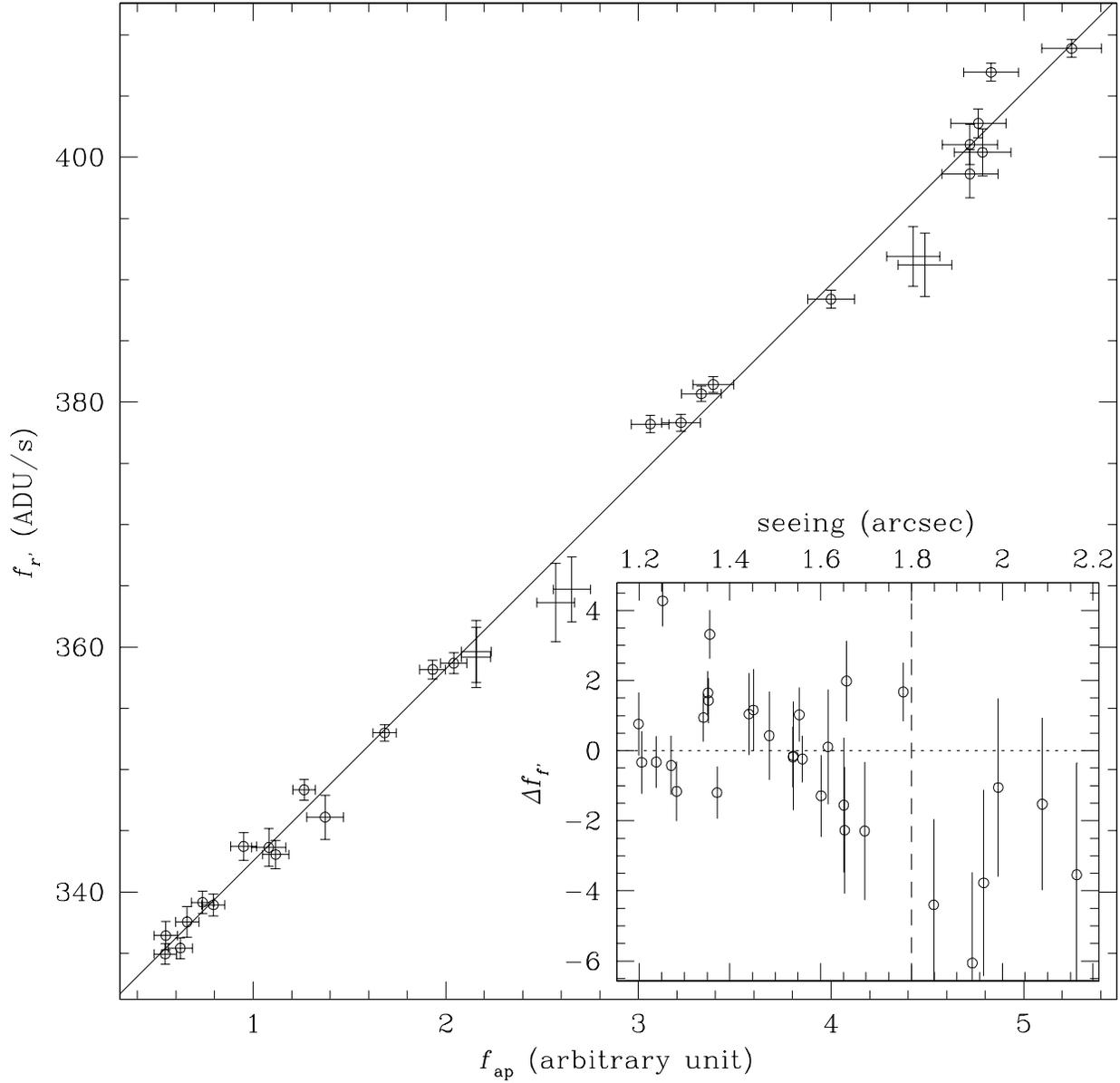}
\caption{\label{apversusp}
Superpixel flux in $r'$ band plotted against the fixed-aperture
photometry result. Error bars without points corresponds to the
datapoints with seeing worse than 1\farcs8. The solid line is the best
fit linear regression line. Residuals of superpixel flux from this
regression versus seeing is also shown in inset.
}\end{figure}

One way of testing whether the correlation found in
Figure~\ref{correlations} is purely statistical or betrays a causal
relation is to derive a second lightcurve which makes different
assumptions as to the effects of seeing variation on the
photometry. We perform fixed-aperture relative photometry of PA-99-N2
on the raw $r'$ frames with respect to stable stars nearby. This is
possible for 31 frames near the peak of the event. In fixed-aperture
photometry, the same fraction of flux from the source star enters the
aperture irrespective of the brightness of the star, for given
seeing. The ratio of flux within the aperture between two stars is
therefore the same as the intrinsic ratio, regardless of seeing. The
linear regression shown in Figure~\ref{apversusp} shows consistency
between the superpixel and fixed aperture lightcurves, with no outlier
worse than 2$\sigma$. The residuals versus seeing shown in the inset
suggest there is no correlation when seeing is better than 1\farcs8
although the superpixel flux is lower compared to the fixed aperture
photometry when seeing is worse than 1\farcs8, which corresponds to
the worst outliers at JD$'$ 63 and JD$'$ 69. At outset, we might
expect the outliers to have little influence on any fitting result, as
the errorbars are large. This is confirmed by adjusting the datapoints
with seeing worse than 1\farcs8 to correspond to the fixed-aperture
photometry and checking that any results do not change
qualitatively. In particular, the measured parameters change by less
than the quoted uncertainty. In Figure~\ref{finallightcurve}, we
present the residuals from a standard Paczy\'nski fit of the 
fixed-aperture lightcurve. Again, the discrepancy from the Paczy\'nski
fit is $\sim$10\%, which is larger than the formal error.

The MEGA collaboration \citep{JDJ} have recently published a
lightcurve for the same event, derived from a difference image
analysis. They plot only the daily-averages for epochs of good seeing.
We find that their lightcurve is consistent with the superpixel
lightcurve~\footnote{This was judged from a comparison between our
superpixel data and their data as read from the lightcurve presented
in \citet{JDJ}.}

The seeming consistency of the three methods makes it unlikely that
the deviations are caused by the data reduction. Although it is still
possible that some unknown systematic problem with the data taking
itself could cause such a spurious effect, we have no reason to
suspect this at the moment.

\subsection{Possible Causes}

There are a number of possible origins for the bump in the lightcurve,
including (i) intrinsic variability due to either the source star or
another neighbouring star, (ii) a close approach to a lensing caustic
\citep{Al02}, or (iii) perturbations caused by a binary source
\citep[see also \citealt{Do98}]{GH92,HC97}. Parallax effects can also
cause perturbations \citep[see also \citealt{Sm02};]
[Appendix~\ref{appb}]{Go92,GMB94}, but the timescale of deviations is
typically $\ga0.25\ \mbox{yr}$, much longer than that of the bump in
PA-99-N2. Hence, we discount this possibility.

Plots of the rms variation on each pixel of $r'$ band images taken
during 2000 and 2001 seasons show no evidence of any additional
variable source within the 7$\times$7 superpixel. There is also no
evidence for any variation in the light centroid during the course of
the event, which might be expected if the lightcurve is being polluted
by a nearby variable. This provides good evidence that the bump is not
caused by a nearby variable. It is possible that the source itself is
variable, but this is not particularly favored because the deviations
are achromatic and because the lightcurve during 2000 and 2001 seasons
appears to be flat to good accuracy.

A binary source is somewhat disfavored on the grounds that the
perturbations to the lightcurve are usually almost periodic. However,
this possibility cannot be definitely ruled out.

In the following, we assume that the deviations are caused by a binary
lens. We find that there are physically reasonable binary lens models
that are consistent with the available photometric data of PA-99-N2 to
a high statistical significance.

\section{Binary Lens Model}
\label{model}

The most straightforward way of generating fold caustics is with
binary lenses. The simplest cases of binary lens lightcurves -- a
point source in uniform rectilinear motion lensed by a static binary
lens -- are characterized by six parameters: the ratio of the two
masses $q$, the projected separation in terms of the Einstein ring
radius $d$, the angle at which the source motion crosses the binary
axis $\alpha$, the time $t_0$ and the distance $u_0 = u(t_0)$ of the
closest approach to a fixed reference point\footnote{Following
\citet{Al02}, the reference point is chosen to be either the center of
the mass if $d\leq1$, or the primary-lens position shifted toward the
secondary by the zeroth order correction, $[(1+q^{-1})d]^{-1}$ if
$d\geq1$.} and the Einstein timescale $t_{\rm E}$. The best fit binary
lens lightcurve of PA-99-N2 can be found by searching for the $\chi^2$
minimum in this six-dimensional parameter space using any standard
minimization technique. Note that, while two further parameters for
each passband are required to relate the lensing magnifications to the
observed fluxes, these parameters enter linearly into the fit, and
thus, this part of the problem can be decoupled from the rest of the
problem and solved directly by separate matrix inversion.

\begin{figure}
\plotone{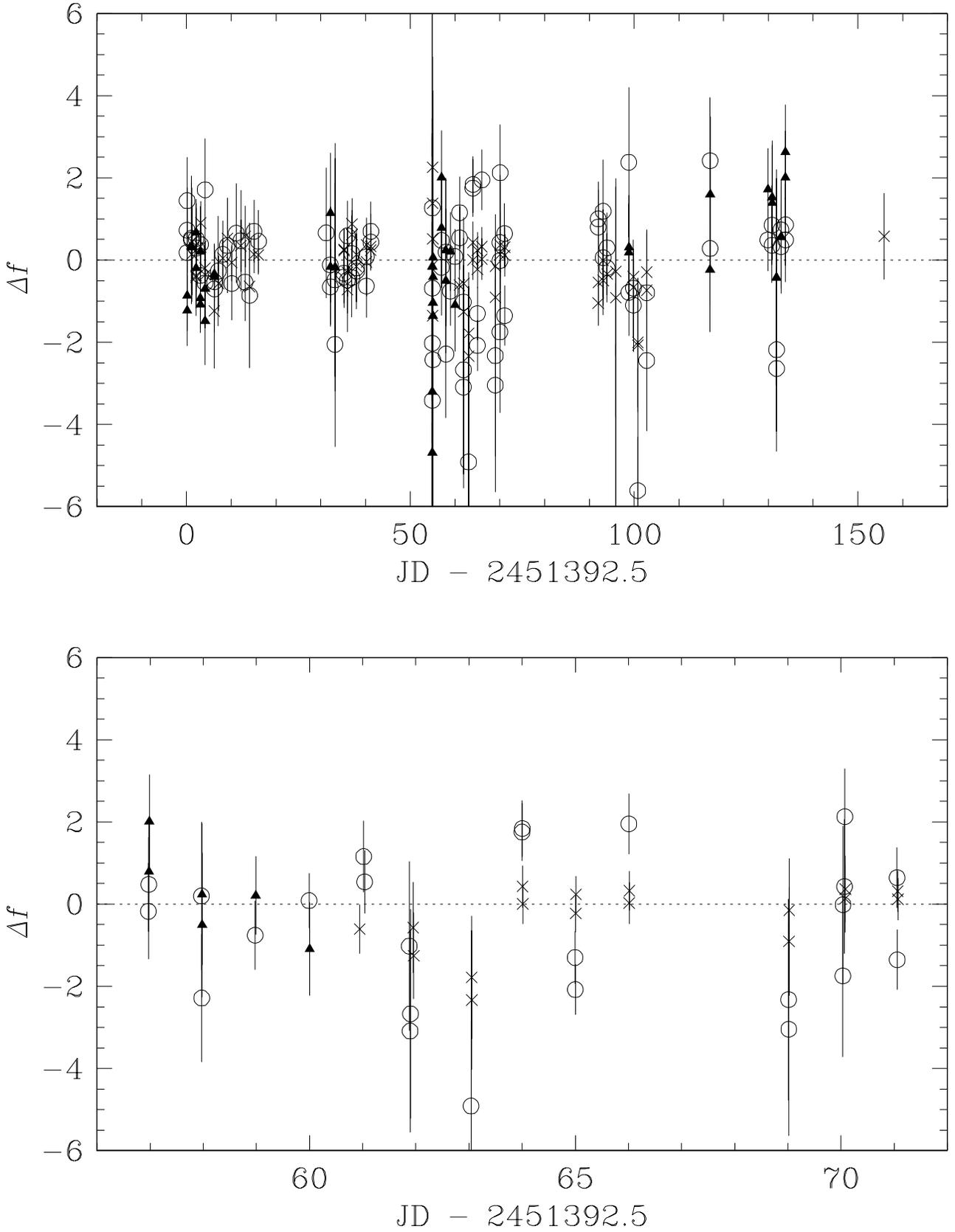}
\caption{\label{binresidualstwo}
Residuals from the best fit binary lens model (C1). Symbols are the
same as Fig.~\ref{binresiduals}. The upper panel is for the whole 1999
season while the lower panel zooms in the rising part of the
lightcurve near the time of the anomaly.
}\end{figure}

We have carried out this minimization with the first two seasons of
data. There are 212 datapoints (65 in $g'$, 102 in $r'$ and 45 in
$i'$) from the 1999 season and 276 datapoints (146 in $r'$ and 130 in
$i'$) from the 2000 season.\footnote{The difference of the quoted
number of datapoints between \citet{PH03} and this paper is because
\citet{PH03} used daily averaged datapoints while here we use a single
point for each nonfaulty frame.} We have found a number of possible
binary lens models, and their parameters are listed in
Table~\ref{binarymodel}. For the formally best fit model (C1), the
lightcurve residuals are shown in Figure~\ref{binresidualstwo}. We
find that there exist several binary lens models that lie at a local
$\chi^2$ minimum for which the $\chi^2$ difference with respect to the
global minimum is quite small. In particular, the two best fit models,
C1 and W1, are essentially indistinguishable in terms of their values
of $\chi^2$. Some of these degeneracies would have been broken if the
$\sim$20-day observation gap between JD$'$ 72 and JD$'$ 91 had been
regularly covered (for example, Model C2 predicts a cusp grazing near
JD$'$ 74 and Model W1$'$ shows a second bump around JD$'$ 77) while
others are much more difficult to discriminate.

From Table~\ref{binarymodel}, the timescale of the two best fit binary
models are $125.0\pm7.2\ \mbox{days}$ (C1) and $132.3\pm7.2\
\mbox{days}$ (W1). In either case, the timescale is larger than
$\sim$92 days reported by \citet{PH03} under the assumption of a
single point mass lens. This may be expected from the fact that the
binary fit possesses a higher maximum amplification. Examination of
images taken well after the event shows no resolved star in the
vicinity of PA-99-N2. On the other hand, a barely resolved nearby star
at this approximate location in our $r'$ band image is measured to
have a flux of 7 ADU s$^{-1}$ \citep{PHthesis}. However, the best
estimate of the $r'$ band source flux of PA-99-N2 from the Paczy\'nski
fit is $7.90\pm0.42\ \mbox{ADU s$^{-1}$}$ and thus nominally
inconsistent with it being unresolved at baseline at the 2$\sigma$
level. This problem is ameliorated with the binary fit, as the source
is predicted to be fainter, namely $4.76\pm0.34\ \mbox{ADU s$^{-1}$}$
(C1) or $4.55\pm0.33\ \mbox{ADU s$^{-1}$}$ (W1), well below the
detection limit of a resolved star.

Using the transformation of \citet{PHthesis}, the baseline source flux
is calculated as $V=23.44\pm0.09$, which is the average of the results
for the two best fit models. The source color is determined from the
achromaticity of the lightcurve -- the slope of the linear regression
line in Figure~\ref{regression} is directly related to the
instrumental $g'_{\rm I}-r'_{\rm I}$ color, which, in turn, can be
translated into $V-R=0.983\pm0.013$. Similarly, the linear regression
involving $i'$ band flux provides the second color
$V-I=2.233\pm0.020$. These colors are roughly consistent with that of
an early M giant \citep{BB88} with $T_{\rm eff}\simeq 3700\ \mbox{K}$
and $\log g\simeq 1$ \citep{LCB97}, although secure identification is
subject to uncertainties due to unknown reddening, metallicity, and
mass of the source. In principle, one can estimate the reddening from
two color measurements, provided that we know the intrinsic relation
between the two colors and the proper reddening law. Here, if we
assume that the local M giants sample of \citet{Fl94} defines an
intrinsic $(V-R)$-$(V-I)$ relation and that the redding law is
described by \citet*{CCM89} with $R_V=3.1$, then we get the
extinction, $A_V=0.54\pm0.21$ and the dereddened colors of the source,
$(V-R)_0=0.85\pm0.05$ and $(V-I)_0=1.95\pm0.11$. While the estimated
extinction appears to be rather large, it is not unreasonable
considering that the Galactic foreground extinction towards M31 is
estimated to be $A_V=0.2\sim0.25$ \citep*{BH82,SFD98}. With the
distance modulus of M31 $(m-M)_{\rm M31}=24.47$ \citep{Ho98,SG98}, the
absolute magnitude of the source is then $M_V\sim-1.6$, which,
combined with its intrinsic color, locates the source on the
color-magnitude diagram as a member of the asymptotic giant branch or
the red giant branch of metal-rich populations. This suggests that the
source probably lies in the M31 disk rather than the metal-poor
spheroid.

The color and the magnitude of the source can be combined to provide
an estimate of the angular source radius $\theta_*$. This can be done
using an empirical calibration of the color with the surface
brightness, such as that of \citet{vB99}. For the source star of
PA-99-N2, the calibration of \citet{vB99}, combined with the
$(V-I)$-$(V-K)$ correlations of \citet{BB88}, provides the measurement
of its angular radius $\theta_*=0.51\pm0.06\ \mbox{$\mu$as}$. We note
that the result is insensitive to the extinction estimate unless the
reddening law through the line of sight is abnormal. That is,
\begin{equation}
\frac{(\delta\theta_*)_{\delta A_V}}{\theta_*}={\ln 10 \over 5}
\left[1-1.115\,\frac{d(V-K)}{d(V-I)}\frac{E(V-I)}{A_V}\right]\,
\delta A_V,
\end{equation}
and so, with $d(V-K)/d(V-I)\simeq1.67$ \citep{BB88} and $R_{VI}=1.92$
\citep{CCM89}, even if $A_V$ were mismeasured by $\sim$0.4 mag, the
resultant error in $\theta_*$ would be only $\sim$0.5\%.\footnote{The
effect of extinction is somewhat larger if the reddening law follows
$R_{VI}=2.41$ \citep{SFD98}, that is, $\delta\ln\theta_*/\delta A_V
\simeq0.1$.} We also check the result with a different calibration
based on $V-I$ \citep{BBH99}, and get a consistent answer
$\theta_*=0.51\pm0.05\ \mbox{$\mu$as}$. The corresponding linear
radius of the source is $\sim$85 R$_\sun$, and the implied surface
gravity
\begin{equation}
\log \left(g\over\mbox{cm s$^{-2}$}\right)=
0.566+\log\left({\mathcal{M}\over\msun}\right)
-2\log\left({\theta_*\over0.51\ \mbox{$\mu$as}}\right)
-0.4\left[(m-M)_{\rm M31}-24.47\right],
\end{equation}
is also consistent with the source being a giant. There is some
evidence for the presence of finite source size effects in the
lightcurve and this, combined with the above estimate of the angular
source size, allows us to infer further properties of the source. This
is explored in Appendix~\ref{appa}.

\section{The Nature of the Lens}

The optical depths of different lens populations may be used to assess
the likely nature of the lens. The advantage of using the optical
depths for this purpose is that they are reasonably robust. Analysis
of the timescale or other parameters when available can give further
clues as to the nature of the lens, although this is usually more
model-dependent. Here, we provide the assessment of the most probable
lens population using the optical depth, the observed event duration
and the constraints on the lens-source relative proper motion. Each
level of assessment exploits more of the available information but
requires more model assumptions.

Optical depth evaluations require an assumed spatial distribution for
the lenses and sources. We consider the M31 disk and spheroid to be
plausible locations for the source and for a stellar lens. We consider
massive compact halo objects (MACHOs) in the M31 and the Milky Way
dark haloes to be additional possibilities for the lens. The M31 disk
is taken to have a double-exponential density profile with a scale
length of 5.8 kpc, scale height of 0.4 kpc and central density of
$0.4\ \mbox{\msun\ pc$^{-3}$}$ \citep[e.g.,][]{Hodge92,Ke01}. For the
spheroid, we adopt the flattened power-law distribution of
\citet{RGG98} with a core radius of 1 kpc and central density of $5\
\mbox{\msun\ pc$^{-3}$}$. The M31 and Milky Way haloes are taken to be
near-isothermal spheres as in \citet{Ke01}, with both haloes truncated
at 100 kpc. We assume that the MACHOs comprise 20\% of the total halo
mass \citep{Al00}. The combined mass distribution of the M31
populations provide a good fit to the rotation curve \citep{kent89}
and to the radially-averaged $R$-band surface brightness profile
\citep{wal87} for reasonable stellar mass-to-light ratios.

Table~\ref{opticaldepths} lists the optical depth $\tau$ for the
various combinations of lens and source populations. To compute the
relative likelihood of the lens and source combinations, the relative
density of the spheroid and disk source populations must be taken into
account. At the location of PA-99-N2, the disk surface luminosity
density is 4.8 times larger than that of the spheroid for our assumed
model. (Strictly speaking, the proper ratio of the source density
should be that of the specific surface density constrained by the
observed source color and magnitude. Due to the lack of a proper local
population model of the disk and spheroid stars at the position of
PA-99-N2, we ignore the source magnitude and color measurement.)
After factoring in the relative source densities, we find the most
likely scenario for PA-99-N2 is a source star in the M31 disk and a
lens in the M31 halo, (with probability $P = 0.35$) followed by the
disk self lensing ($P = 0.2$) and lensing of a disk star by a spheroid
star ($P = 0.18$). A MACHO is the more likely alternative for the lens
than a star, provided that MACHOs comprise no less than 15\% of the
total halo mass. However, if we accept that the metal-rich nature of
the source implies a denizen of the M31 disk, then the scenario of a
lens in the M31 halo becomes slightly more probable ($P = 0.40$).

To utilize the information on the timescale measurement, further
assumptions are required as to the lens and source velocity
distributions, as well as the lens mass function. We assume random
velocities to be characterised by a Maxwellian distribution with a
one-dimensional dispersion $\sigma=$ 40, 100, 166 and 156 km s$^{-1}$
for the M31 disk, spheroid, halo and Milky Way halo, respectively. In
addition, we include rotational velocity components of 235 and 30 km
s$^{-1}$ for the M31 disk and spheroid, respectively. M31 stars are
assumed to follow the Solar-neighborhood mass function, specifically a
broken power law where $dn/dm\propto m^{-1.4}$ between 0.01 \msun\ and
0.5 \msun\ and $dn/dm\propto m^{-2.2}$ above 0.5 \msun, with
continuity enforced at the knee \citep{GBF97}. For disk stars the
upper mass cut-off is 10 \msun, while for spheroid lenses we assume a
1-\msun\ cut-off. The MACHO mass function is assumed to be a Dirac
delta function at 0.5 \msun\ \citep{Al00}.

It should be stressed that, while we are not in the classical
microlensing regime, we can use the classical microlensing rate to
compare relative likelihoods. This is because the Einstein timescale
is uniquely specified from the fit, and so the pixel-lensing detection
efficiency is the same unknown constant for all potential lens
populations and can therefore cancels out in any comparison.
Similarly, the source magnitude is also specified by the fit, and we
therefore do not need to make any assumptions regarding the M31
stellar luminosity function.

Table~\ref{opticaldepths} then shows the differential microlensing
rate per source, $d\R/dt_{\rm E}$, evaluated at $t_{\rm E,0}=
125.0\pm7.2\ \mbox{days}$. The most likely scenario for this case is
an M31 MACHO lensing a disk star ($P = 0.41$), followed by a disk
source lensed by a Milky Way MACHO ($P = 0.27$) and then disk self
lensing ($P = 0.11$). While this result is broadly consistent with
those derived from the optical depth values, the observed timescale
lends greater preference to MACHOs as a probable lens population for
our assumed model. Note too that if we use differential rates to make
the judgement on the origin of the lens, then a MACHO is more likely
than a stellar lens provided MACHOs comprise more than 5\% of the M31
and Milky Way haloes.

\begin{figure}
\plotone{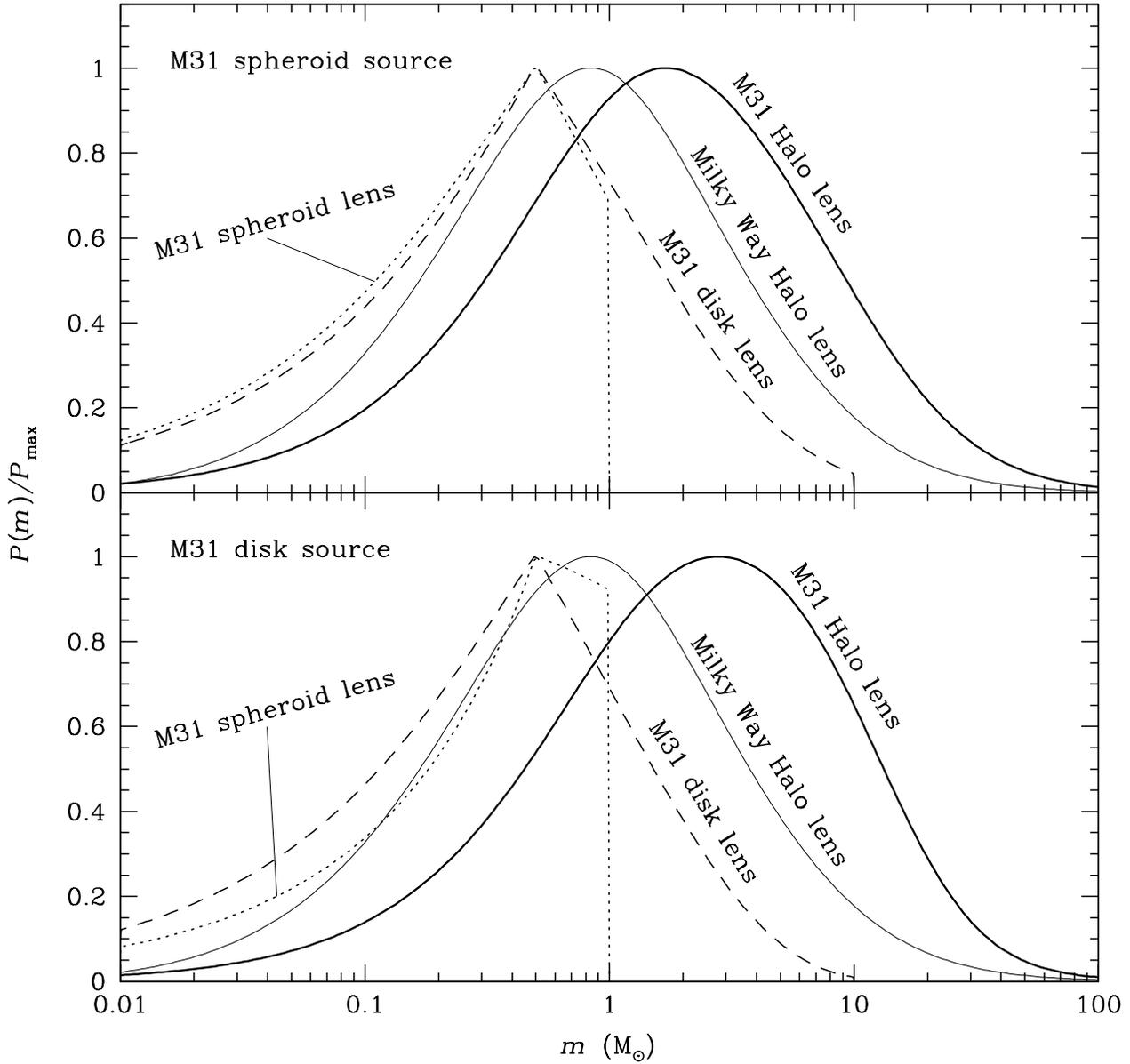}
\caption{\label{massestimate}
Relative probability $P$ (or differential microlensing rate $d^2\R/(dm
dt_{\rm E})$) for the lens mass $m$ for different lens locations. The
upper (lower) panel assumes that the source lies in the M31 spheroid
(or M31 disk). For the halo lenses, these curves show the variation of
the differential microlensing rate, $d\R/dt_{\rm E}$ when the halo is
comprised of objects of the given mass in entirety. All curves are
normalised to the same maximum value $P_{\rm max}$.
}\end{figure}

Figure~\ref{massestimate} shows the posterior probability for the lens
mass for each combination of lens and source when the timescale
information is considered. Here, for the case of MACHO lenses, we
relax the assumption of a MACHO being 0.5 \msun. But, MACHOs are still
assumed to be a population with a unique characteristic mass. If the
lens is a MACHO in the Milky Way halo, the corresponding 95\%
confidence interval on its mass is 0.04-13 \msun. Similarly, if a
compact object in the M31 halo is responsible for the event, its mass
is predicted to lie between 0.09-32 \msun\ (95\% confidence; the disk
source) or 0.06-30 \msun\ (95\% confidence; the spheroid source).
While both of these mass ranges are quite broad, they tend to favor an
interpretation of the MACHO as a stellar remnant or primordial black
hole. On the other hand, if PA-99-N2 is a disk self-lensing event, the
most probable lens mass is 0.5 \msun (which is at the break of the
mass function), and the associated 95\% confidence interval is
0.02-3.6 \msun. In this case, the mass of the primary is consistent
with it being a typical low mass disk star, and the mass of the
secondary lies below the hydrogen-burning limit at better than the
95\% confidence limit. The case of a spheroid source and a disk lens
leads to a similar mass estimate (0.02-4.7 \msun), while a spheroid
lens results in a somewhat narrower mass range (0.02-1 \msun) due to
the lowering of the upper mass cut-off compared with the disk mass
function.

\section{Conclusions}

We have studied the anomaly in the lightcurve of PA-99-N2. Although
there are other possibilities, the assumption of a binary lens
provides the most economical explanation of the data. The lens may lie
either in the M31 disk/spheroid or the Milky Way/M31 haloes. If the
lens lies in the disk/spheroid, the primary is a low mass star. Its
companion lies below the hydrogen burning limit with 95\% confidence.
This would makes it the most distant brown dwarf so far discovered. If
the lens lies in the halo, then the primary is most probably a stellar
remnant and its companion either a brown dwarf or low mass star.

\acknowledgments Work by JA has been supported through a grant from the
Leverhume Trust Foundation. NWE acknowledges financial support from
the Royal Society (UK). SCN was supported by the Swiss National Science
Foundation and by the Tomalla Foundation. Work by AG was supported by
the grant AST 02-01266 from the National Science Foundation (US).

\appendix

\section{The Finite Source Effect and the Proper Motion Constraint}
\label{appa}

When a caustic passes close to a source star, the point-source
approximation is sometimes no longer valid. If this ``finite source''
effect is large enough, the lightcurve can be analyzed to derive an
additional parameter $\rho_*=\theta_*/\theta_{\rm E}$, where
$\theta_{\rm E}$ is the angular Einstein ring radius. However, we see
that the point-source approximation is already good for the lightcurve
of PA-99-N2, and so the finite source effect is not expected to be
detected with a high statistical significance. Hence, we do not
perform a full search for the best fit finite-source binary-lens model
in the seven-dimensional parameter space. Instead, we check whether
the deviations of finite-source lightcurves near the best fit
binary-lens model are consistent with the observations.

In fact, we find a marginal improvement of the fit by incorporating
the finite source effect. A restricted search around the best fit
point-source model C1 yields a model (FS in Tab.~\ref{binarymodel})
with a smaller $\chi^2$ than that of C1 by 4.9. We note that, while
the geometry of the binary lens -- $d$ and $q$ -- has been driven to
move along the principal conjugate direction (towards the smaller
$q$), the measurements of most other parameters are only minimally
affected by the incorporation of the finite source effect. In
particular, the timescale and the source flux are changed only by $\la
0.5\%$, which is an order of magnitude smaller than their associated
uncertainties. The limiting value when the finite source fit is no
better than the point source fit is $\rho_*<2.06\times 10^{-2}$
(formally corresponding to $\Delta\chi^2<4.9$). Combining this with
the measurement of $t_{\rm E}$ and $\theta_*$ (see \S~\ref{model})
provides us with the constraint on the angular Einstein ring size
$\theta_{\rm E}$ and the lens-source relative proper motion 
$\mu_{\rm rel}$;
\begin{mathletters}
\begin{equation}
\theta_{\rm E}\,(=\theta_*\rho_*^{-1})\,\ga25\ \mbox{$\mu$as}
\end{equation}
\begin{equation}
\mu_{\rm rel}\,(=\theta_{\rm E}t_{\rm E}^{-1})\,\ga
0.20\ \mbox{$\mu$as day$^{-1}$}
=3.4\times 10^{2}\ \mbox{km s$^{-1}$ Mpc$^{-1}$}.
\end{equation}
\label{eq:cut}
\end{mathletters}
Caution should be used in interpreting these limits, as the
uncertainties may not properly include systematic effects such as
uncertainty in the color transformations or any possible problems in
the photometry for the critical three nights data over which finite
source effects are detected. For this reason, we distinguish these
from other more robust results and have first presented an analysis
without any constraint on the proper motion. Nonetheless, we note
that, if we combine the limit on the angular Einstein radius with the
distance to M31, the implied separation of the two binary lens
components (for M31 lenses) is $\ga 10\ \mbox{AU}$.

We incorporate the lens-source relative proper motion information into
the assessment of the nature of the lens by deriving the differential
microlensing rate per source, $d\R/(dt_{\rm E} d\mu_{\rm rel})$,
averaged over the probability distribution of $\mu_{\rm rel}$ inferred
from the finite source effect on the observed lightcurve.  Here, the
differential microlensing rate per source can be found for the case of
Maxwellian lens and source random velocity distributions by
\begin{equation}
\frac{d^2\R}{dt_{\rm E}\,d\mu_{\rm rel}}=
\int\int\frac{4\mu_{\rm rel}^3}{\sigma_{\rm T}^2\Sigma_{\rm S}} 
\exp\left(-\frac{v_{\rm S}^2}{2\sigma_{\rm T}^2}\right) 
\exp \left(-\frac{\mu_{\rm rel}^2D_{\rm L}^2}{2\sigma_{\rm T}^2}\right)
I_0\left(\frac{\mu_{\rm rel}D_{\rm L}v_{\rm S}}{\sigma_{\rm T}^2}\right) 
\frac{D_{\rm L}^7}{D_{\rm S}(D_{\rm S}-D_{\rm L})}
\rho_{\rm S}(D_{\rm S})\rho_{\rm L}(D_{\rm L})\psi(m)dD_{\rm S}dm,
\label{drdtdmu}
\end{equation}
where $D_{\rm S}$ is the distance to the source, $\rho_{\rm L}$ and
$\rho_{\rm S}$ are the lens and source mass densities, 
$\Sigma_{\rm S}$ is the source surface mass density, $\psi(m)=
\rho_{\rm L}^{-1}(dn/dm)$ is the lens mass function normalized to the
lens mass density, and $I_0(y)$ is the Modified Bessel function of
order zero. Also note that the distance to the lens, $D_{\rm L}$ is in
fact a function of $D_{\rm S}$ and the lens mass $m$ for a fixed
$t_{\rm E}$ and $\mu_{\rm rel}$. Equation~(\ref{drdtdmu}) has been
derived by attributing all random velocities to the lens and all
rotational motions to the source. The velocity dispersion 
$\sigma_{\rm T}$ is therefore a combination of the lens velocity
dispersion $\sigma_{\rm L}$ and source velocity dispersion
$\sigma_{\rm S}$ projected at the lens location, while $v_{\rm S}$ is
the transverse component of the difference between source and lens
rotation speeds. For Milky Way lenses $\sigma_{\rm T}\simeq 
\sigma_{\rm L}$, while for M31 lenses $\sigma_{\rm T}\simeq
(\sigma_{\rm L}^2+\sigma_{\rm S}^2)^{1/2}$.

In Table~\ref{opticaldepths}, we list this averaged differential rate
for the same various combinations of the lens and source populations.
Note that the absolute normalization of the probability density for
$\mu_{\rm rel}$ is unknown, and so we have effectively integrated
$d^2\R/(dt_{\rm E} d\mu_{\rm rel})$ over $\mu_{\rm rel}$ weighted by
its relative probability. Hence, the numerical values of the
differential rates are essentially normalized by the peak value of the
probability density for $\mu_{\rm rel}$. We find that the derived
proper motion constraint overwhelmingly favors an M31 MACHO lens with
either a disk or spheroid star being a source. In fact, an M31 halo
with a MACHO fraction as small as 1\% produces a comparable event rate
as that of the stellar lenses.

\section{Seeing Correction Plus Parallax Effects}
\label{appb}

We have already shown in \S~\ref{data} that the bump in the lightcurve
of PA-99-N2 can be removed by correlated variations with the seeing,
but that the resulting lightcurve is still anomalous. Here, we explore
the possibility that the remaining anomaly is due to some physical
effect. An asymmetry between the rising and falling side of the
lightcurve is characteristic of a non-uniform lens motion with respect
to the source. The most common cause of this is the Earth's orbital
motion around the Sun, referred to as the parallax effect.

We fit the data of PA-99-N2 to equation~(\ref{eq:seevar}) with
$f_{\rm true}$ as the lightcurve of a parallax event. The deviation is
consistent with the parallax effect and there are at least four
degenerate solutions, listed in Tables~\ref{parallax} and \ref{par2}.
(This arises because it is only an acceleration with respect to the
instantaneous relative velocity of the lens that is detected.)
However, we regard this as a less likely alternative than the
hypothesis of a binary lens, even more so than suggested on purely
statistical grounds by the $\chi^2$ difference.

First, using the data in Table~\ref{par2}, the projected velocity of
the lens on the observer plane ($\tilde v$) is $\sim$20 km s$^{-1}$.
We note that this is an upper limit to the lens-source relative
velocity. Already this makes it extremely unlikely that the lens
resides in the haloes of either the Milky Way or M31. Only Milky Way
disk stars are likely to have such a small relative velocity. The
distance to the lens belonging to the Milky Way can be estimated from
the parallax, assuming $D_{\rm L}\ll D_{\rm S}$, that is,
\begin{equation}
D_{\rm L}\approx 
D_{\rm L}\left(1-{D_{\rm L}\over D_{\rm S}}\right)^{-1}=
{c^2\tilde r_{\rm E}^2\over 4G\mathcal{M}}=
0.1227\ \mbox{kpc}\ \left({\mathcal{M}\over\msun}\right)^{-1}
\left({\tilde r_{\rm E}\over\mbox{AU}}\right)^2.
\end{equation}
For a typical lens mass of $\sim$0.5 \msun, then the lens distance is
$\sim$2 kpc. This corresponds to a vertical height above the Galactic
plane of $\sim$700 pc. This is uncomfortably high, as it is twice the
scale-height of the thin disk. The location of the lens can be brought
within the scaleheight of the thin disk by increasing its mass to
$\sim$1 \msun, but such an object would surely be directly imaged if
it were a normal star. The only possibility that seemingly remains is
a disk white dwarf.

\begin{figure}
\plotone{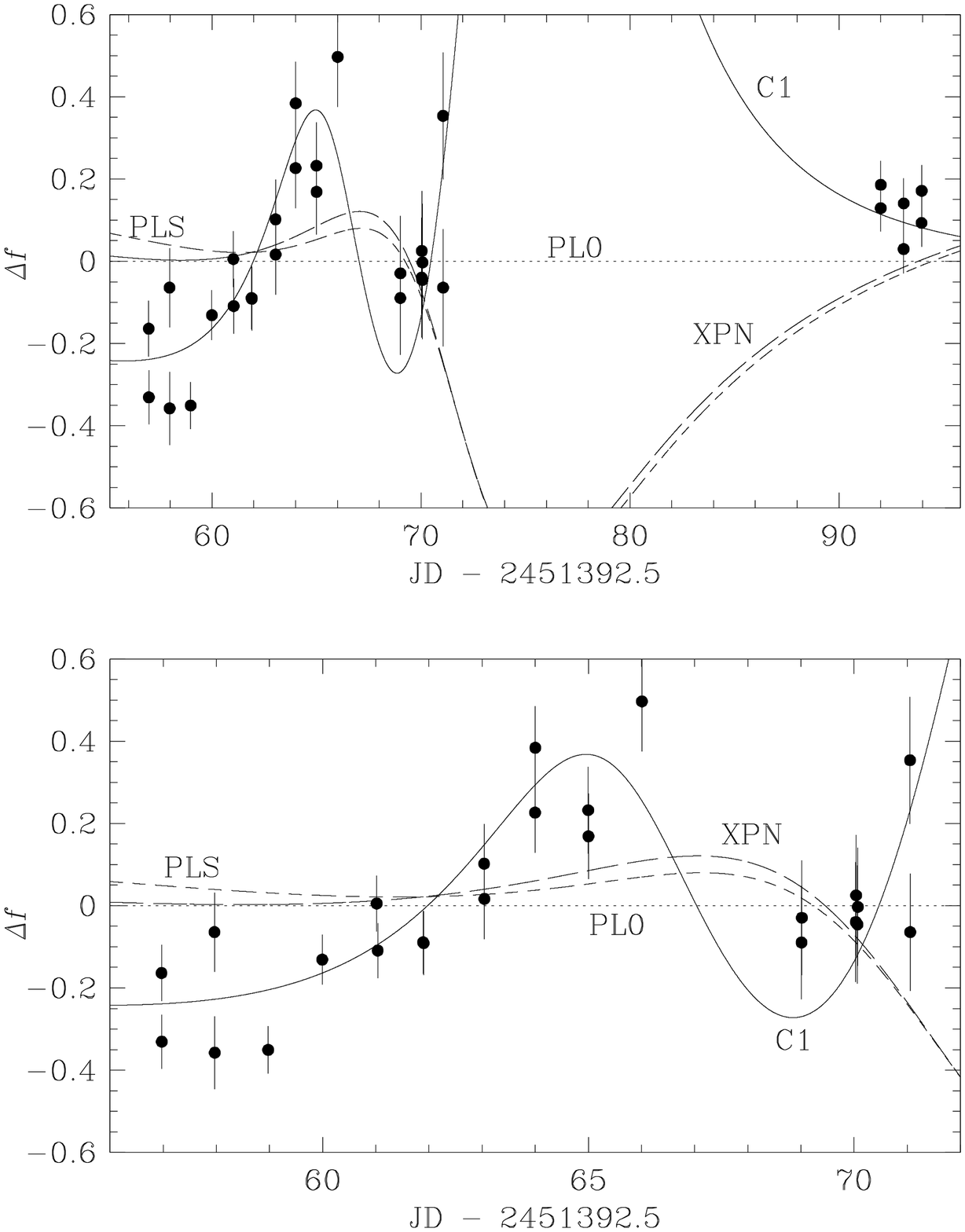}
\caption{\label{finallightcurve}
Residuals of the fixed-aperture photometry from a standard Paczy\'nski
fit. 
Also shown are a
number of models discussed in the text: Paczy\'nski curve with
seeing correction (PLS), best fit binary lens model (C1) and best fit
parallax model with seeing correction (XPN). For the details, refer to
Tables~\ref{binarymodel} and \ref{parallax}.
}\end{figure}

In addition to the \emph{a priori} low probability of lensing by a
disk white dwarf, the further objection to this scenario -- seeing
variation plus parallax effect models -- is that it is not consistent
with the fixed aperture photometry described in \S~\ref{data}. This is
evident on inspection of Figure~\ref{finallightcurve}, which shows
residuals of the fixed aperture photometry together with a number of
the models discussed in the text. The best fit binary lens model (C1)
clearly describes best the pattern of deviations among various
alternatives.


\begin{deluxetable}{lrrcccccc}
\tablewidth{0pt}
\tablecaption{\label{binarymodel}
Binary Lens Models of PA-99-N2}
\tablehead{
\colhead{model}&
\colhead{$\log d$}&
\colhead{$\log q$}&
\colhead{$t_0$}&
\colhead{$u_0$}&
\colhead{$t_{\rm E}$}&
\colhead{$\alpha$ \tablenotemark{a}}&
\colhead{$\chi^2$}&
\colhead{$\chi^2_{1999}$ \tablenotemark{b}}\\
&&&(JD $-$ 2451392.5)&($\times10^{-2}$)&(days)&(deg)&&}
\startdata
C1&
$-0.242\pm0.020$&
$-1.850\pm0.096$&
$73.22\pm0.19$&
$3.60\pm0.37$&
$125.0\pm7.2$&
\phn$24.2\pm2.1$&
312.5&195.3\\
W1&
$0.265\pm0.021$&
$-1.911\pm0.111$&
$73.52\pm0.25$&
$3.40\pm0.40$&
$132.3\pm7.2$&
\phn$24.5\pm2.7$&
312.7&195.6\\
W1$'$&
$0.126\pm0.021$&
$-2.57\pm0.20$\phn&
$75.84\pm0.10$&
$0.66\pm0.23$&
$172.\pm23.$\phn&
\phn\phn$4.9\pm1.0$&
322.3&205.7\\
C2&
$-0.581\pm0.023$&
$-0.384\pm0.098$&
$73.18\pm0.10$&
$2.10\pm0.49$&
$129.0\pm7.7$&
$271.5\pm4.7$&
345.6&228.8\\
C3&
$-0.433\pm0.049$&
$-0.985\pm0.156$&
$72.73\pm0.12$&
$4.09\pm0.24$&
$121.6\pm6.6$&
$278.1\pm3.6$&
345.7&228.9\\
W2&
$0.78\pm0.10$\phn&
$0.03\pm0.24$\phn&
$73.17\pm0.11$&
$1.44\pm0.32$&
$195.\pm28.$\phn&
$270.7\pm4.9$&
346.5&229.8\\
W3&
$0.46\pm0.10$\phn&
$-0.93\pm0.27$\phn&
$72.72\pm0.13$&
$3.98\pm0.23$&
$130.8\pm6.2$&
$278.5\pm5.4$&
349.7&232.9
\\[1ex]
FS \tablenotemark{c}&
$-0.170\pm0.047$&
$-2.12\pm0.20$\phn&
$73.12\pm0.13$&
$3.86\pm0.51$&
$124.3\pm8.5$&
\phn$26.4\pm2.7$&
307.6&190.3\\
PL \tablenotemark{d}&
\nodata&\nodata&
$71.56\pm0.08$&
$7.47\pm0.37$&
\phn$91.9\pm3.7$&
\nodata&
824.6&700.8
\enddata
\tablenotetext{a}
{The fixed reference point of system lies on the right hand side of
the moving source path.}
\tablenotetext{b}
{The $\chi^2$ only accounting for the 1999 season data.}
\tablenotetext{c}
{The best fit finite source model associated with C1. The finite
source parameter for this model is derived to be $\rho_*=
(1.31\pm0.30)\times 10^{-2}$. One may formally interpret that C1 is in
fact the best fit model with the constraint that $\rho_*=0$.}
\tablenotetext{d}
{The best fit Paczy\'nski lightcurve. The difference of the errorbars
of the parameter measurement and $\chi^2$ compared to \citet{PH03} is
because here we use individual datapoints (instead of daily averaged
points) and the formal errorbars derived from the local curvature
matrix.}
\tablecomments
{The total number of degrees of freedom is 476 ($=488\ \mbox{datapoints}
-12\ \mbox{parameters}$), except PL where nine parameters are used.
The uncertainties for parameters are derived from the local curvature
matrix (half the local Hessian of $\chi^2$) except for model W1$'$,
for which they represent the interval of $\chi^2\leq\chi^2_0+1$ where
$\chi^2_0$ is the associated local minimum.}
\end{deluxetable}

\begin{deluxetable}{llccc}
\tablewidth{0pt}
\tablecaption{\label{opticaldepths}
Microlensing Optical Depths and Event Rates at the Location of
PA-99-N2} 
\tablehead{
\colhead{lens}&
\colhead{source}&
\colhead{$\tau$}&
\colhead{$d\R/dt_{\rm E}$}&
\colhead{$\langle d\R/(dt_{\rm E} d\mu_{\rm rel}) \rangle$}\\
&&\colhead{($\times 10^{-7}$)}&
\colhead{($\times 10^{-9}$ yr$^{-1}$ day$^{-1}$ star$^{-1}$)}&
\colhead{($\times 10^{-10}$ $P_{\mu_{\rm rel},\max}$
yr$^{-1}$ day$^{-1}$ star$^{-1}$)}}
\startdata
M31 disk&M31 disk&
2.8&0.59&0.25\\
M31 spheroid&&
2.5&0.35&0.03\\
M31 halo&&
4.8&2.2&4.5\\
Milky Way halo&&
2.0&1.5&0\\[1em]
M31 disk&M31 spheroid&
1.1&0.22&0.35\\
M31 spheroid&&
1.1&0.30&0.02\\
M31 halo&&
3.1&1.8&3.6\\
Milky Way halo&&
2.0&1.5&0
\enddata
\tablecomments
{For both haloes of M31 and the Milky Way, it is assumed that 0.5
\msun\ MACHOs comprise 20\% of the total mass of dark haloes
\citep{Al00}.} 
\end{deluxetable}

\begin{deluxetable}{lcccccl}
\tablewidth{0pt}
\tablecaption{\label{parallax}
Models of PA-99-N2 with a seeing correlation}
\tablehead{
\colhead{model}&
\colhead{$t_{0,\earth}$}&
\colhead{$u_{0,\earth}$}&
\colhead{$t_{{\rm E},\earth}$}&
\colhead{$\chi^2$}&
\colhead{$\chi^2_{1999}$}&
\colhead{Note.}\\
&(JD $-$ 2451392.5)&($\times10^{-2}$)&
(days)&&&}
\startdata
PL0 \tablenotemark{a}&
$71.56\pm0.08$&
\phn$7.47\pm0.37$&
\phn$91.9\pm3.7$&
824.6&700.8&\nodata\\
PLS \tablenotemark{b}&
$70.39\pm0.09$&
\phn$3.98\pm0.40$&
$148.\pm12.$\phn&
438.1&321.1&\nodata\\
PLA \tablenotemark{c}&
$70.20\pm0.09$&
\phn$2.98\pm0.37$&
$200.\pm22.$/$192.\pm20.$&
342.5&225.5&
$\dot\mu_{\rm E}=(4.0\pm1.0)\times10^{-5}\ \mbox{day$^{-2}$}$
\\[1ex]
XPN&
$70.32\pm0.10$&
\phn$4.42\pm0.63$&
$134.\pm17.$\phn&
337.4&220.6&
$\pi_{\rm E}=(3.0\pm1.1)\times10^{-1}$\\
XPP&
$70.38\pm0.10$&
\phn$4.90\pm0.60$&
$121.\pm13.$\phn&
336.5&221.1&
$\pi_{\rm E}=(5.1\pm0.7)\times10^{-1}$\\
XNN&
$70.29\pm0.10$&
$-4.07\pm0.54$&
$149.\pm17.$\phn&
338.2&221.2&
$\pi_{\rm E}=(2.2\pm0.7)\times10^{-1}$\\
XNP&
$70.34\pm0.10$&
$-3.63\pm0.77$&
$172.\pm34.$\phn&
337.8&223.2&
$\pi_{\rm E}=(3.7\pm0.8)\times10^{-1}$
\enddata
\tablenotetext{a}
{Same as the model PL in Table~\ref{binarymodel}.}
\tablenotetext{b}
{Paczy\'nski curve + seeing correlation from eq.~(\ref{eq:seevar}).}
\tablenotetext{c}
{Point mass lens + uniform acceleration + seeing correlation from
eq.~(\ref{eq:seevar}). Two different values of $t_{\rm E}$ correspond
to two degenerate models discovered by \citet{SMP03}.}
\tablecomments
{The parameters for models with non-uniform motion are basically for
the tangent to the trajectory at the time of the closest approach
between the lens and the source ($t=t_0$). The $\dot\mu_{\rm E}$
denotes the proper acceleration normalised to the angular Einstein
ring size (i.e., $\mu_{\rm E}=\mu_{\rm rel}/\theta_{\rm E}$), while
$\pi_{\rm E}$ is the lens-source relative parallax normalised by 
$\theta_{\rm E}$.}
\end{deluxetable}

\begin{deluxetable}{lcccccc}
\tablewidth{0pt}
\tablecaption{\label{par2}
Standard parameterization of parallax models}
\tablehead{
\colhead{model}&
\colhead{$t_{0,\sun}$}&
\colhead{$u_{0,\sun}$}&
\colhead{$t_{{\rm E},\sun}$}&
\colhead{$\tilde r_{\rm E}$}&
\colhead{$\psi$}&
\colhead{$\tilde v$}\\
&(JD $-$ 2451392.5)&($\times 10^{-2}$)&
(days)&(AU)&(deg)&(km s$^{-1}$)}
\startdata
XPN&
\phn$96.2\pm13.2$&
$-1.6\pm2.0$&
$193.\pm34.$&
$3.3\pm1.2$&
$-87.3\pm4.4$&
$29.9\pm15.1$\\
XPP&
$129.4\pm\phn6.5$&
\phn$3.5\pm3.5$&
$246.\pm31.$&
$2.0\pm0.3$&
$-65.6\pm8.0$&
$13.9\pm1.6$\\
XNN&
\phn$96.7\pm10.4$&
$-7.2\pm2.6$&
$189.\pm30.$&
$4.6\pm1.4$&
$-89.0\pm7.5$&
$42.3\pm17.7$\\
XNP&
$157.8\pm\phn5.8$&
\phn$3.2\pm2.2$&
$358.\pm85.$&
$2.7\pm0.6$&
$-61.8\pm4.4$&
$13.1\pm0.9$
\enddata
\tablecomments
{This gives the 5 parameters required to define a parallax lightcurve
using the conventional notation found in \citet{So01}. The projected
velocity $\tilde v=\mu_{\rm rel}(D_{\rm L}^{-1}-D_{\rm S}^{-1})^{-1}=
v_{\rm rel}[1-(D_{\rm L}/D_{\rm S})]^{-1}$ is found by
$\tilde r_{\rm E}/t_{{\rm E},\sun}$.} 
\end{deluxetable}
\clearpage

\newpage

\end{document}